\documentclass[twocolumn,showpacs,preprintnumbers,amsmath,amssymb]{revtex4}

\usepackage{epsfig}
\usepackage{epsf}
\usepackage{graphics}
\usepackage{graphicx}

\begin{document}

\newcommand{\bc}{\begin{center}}
\newcommand{\ec}{\end{center}}
\newcommand{\be}{\begin{equation}}
\newcommand{\ee}{\end{equation}}
\newcommand{\bq}{\begin{eqnarray}}
\newcommand{\ikl}{\int_k}
\newcommand{\eq}{\end{eqnarray}}
\newcommand{\PLB}{{\it{Phys. Lett. {\bf{B}}}}}
\newcommand{\NPB}{{\it{Nucl. Phys. {\bf{B}}}}}
\newcommand{\PRD}{{\it{Phys. Rev. {\bf{D}}}}}
\newcommand{\AOP}{{\it{Ann. Phys. }}}
\newcommand{\PRL}{{\it{Phys. Rev. Lett. }}}
\newcommand{\MPL}{{\it{Mod. Phys. Lett. }}}
\newcommand{\dena}{(k^2-m^2)}
\newcommand{\dd}{\frac{d^2k}{(2 \pi)^2}}

\title{Arbitrary parameters in implicit regularization and democracy within perturbative description of 2-dimensional gravitational anomalies}
\date{\today}

\author{Leonardo A. M. Souza} \email [Corresponding author: ]{lamsouza@fisica.ufmg.br}
\author{Marcos Sampaio} \email[]{msampaio@fisica.ufmg.br}
\author{M. C. Nemes}\email[]{carolina@fisica.ufmg.br}

\affiliation{Federal University of Minas Gerais -
Physics Department - ICEx \\
P.O. BOX 702, 30.161-970, Belo Horizonte MG - Brazil\\
Fax number: +55-31-3499-5600}

\begin{abstract}
\noindent

We show that the Implicit Regularization Technique is useful to display quantum symmetry breaking
in a complete regularization independent fashion. Arbitrary parameters are expressed by finite differences
between integrals of the same superficial degree of divergence whose value is fixed on physical grounds (symmetry requirements or phenomenology). We study Weyl fermions on a classical gravitational background in two dimensions and show that, assuming Lorentz symmetry, the Weyl and Einstein Ward identities reduce to a set of algebraic equations for
the arbitrary parameters which allows us to study the Ward identities on equal footing. We conclude in a renormalization independent way that the axial part of the Einstein Ward identity is always violated. Moreover whereas we can preserve the pure tensor part of the Einstein Ward identity at the expense of violating the Weyl Ward identities we may as well
violate the former and preserve the latter.

\end{abstract}

\maketitle

\section{Introduction}

Quantum mechanical symmetry breakings or anomalies  are important mechanisms in the description of nature. In the theory of eletroweak interactions the cancellation of anomalies when gauge fields couple to currents requires equal numbers of quarks and leptons with charges taking precisely the values of the Standard Model \cite{SM}. On the other hand the existence of certain anomalies is also important. The well-known axial-vector triangle anomaly is crucial to account for the decay of the neutral pi-meson.
Also the quantum breaking of scale and conformal invariance in quantum field theory translated by the Gell-Mann-Low renormalization group explains the diversity of particles in nature. The importance of anomalies pervades different areas in physics entering into the field of string theory \cite{ST}, condensed matter and gravity (leading to an interplay between Physics and Mathematics through topology \cite{BERTLMANN-B}) and supersymmetry
\cite{SUSYAN}.

In many situations the diagrammatic approach (using either dispersion relations or a definite regularization framework) is the best tool to explore quantum symmetry breaking. That is  because general (non-perturbative) statements are usually hard to obtain. Moreover general theorems state that certain type of anomalies are one loop exact \cite{ADLER}.
Although dimensional regularization (DR) is the natural framework for computing Feynman
diagrams in gauge field theories,
the regularization of dimension specific quantum field theories
such as chiral, topological and supersymmetric gauge  theories is more involved. That is because the analytical
continuation on the space-time dimension of the Levi-Civita tensor is not
well-defined whereas supersymmetry is intrinsically defined on the physical
dimension of the model. Such shortcomings may give to spurious
anomalies  and thus naive DR cannot be used to study quantum mechanical symmetry breaking in dimension specific gauge theories. This is particularly important in deciding whether there are  anomalies in supersymmetric gauge theories
or not. A modification of DR called  dimensional reduction is often used  although their consistency cannot be assured beyond  one loop level \cite{DRED}. Moreover the determination of (super)symmetry restoring counterterms in a anomaly free model is often a tedious task.

Implicit Regularization (IR) is a momentum space setting to perform Feynman
diagram calculations in a regularization independent fashion. Consequently IR turns out particularly adequate to
unravel anomalies within  perturbation theory. In IR, the  Lagrangian of
the underlying quantum field theory is not modified because neither an explicit
regulator is introduced nor the dimensionality of the space time needs to be moved away
from its physical dimension. Such features are shared between IR and Differential Renormalization (DfR)\cite{DF}. The latter is a coordinate space framework which is based on a prescription that extends product of distributions into a distribution without intermediate regulator or counterterms. This is performed in a minimal sense by expressing a coordinate space amplitude as a derivative of a less singular expression. The derivatives are meant to act formally by parts on test functions, neglecting divergent surface terms. In this process a logarithmic mass scale naturally emerges and plays the role of renormalization group scale.  An alternative  procedure of DfR which avoids introducing unnecessary renormalization constants and automatically preserves gauge symmetry (at least to one loop level) is called constrained DfR. IR, on the other hand, operates in momentum space which is convenient to compute amplitudes with fixed external momenta. Moreover we have at our disposal an all ready library of momentum space integrals and  Feynman parameter techniques. Nonetheless it is not a simple momentum space version of DfR. It can give us new insights in some calculations as well as understand the origin of certain regularization dependent results which can be fully appreciated in the study of anomalies.

The idea behind IR is to display the ultraviolet behaviour of the amplitude in the form of  loop integrals which depend solely on the loop momenta by using an algebraic identity in the integrand of the amplitude .  Strictly speaking we may assume an implicit regulator to manipulate the integrand just as in the BPHZ framework in which Taylor operators act on the integrand.  However an explicit realization of such regulator is not necessary to bring about the physical content of the amplitude as one  needs {\it{not}} compute the divergent integrals within IR. They
may be subtracted  and absorbed in the counterterms exactly as they stand.
In fact the explicit computation of such infinities is  the origin of spurious symmetry
breakings which may contaminate the physics of the underlying model. It is important to observe that  although BPHZ is a very powerful framework which enables
to construct proofs of renormalizability to  all orders, for gauge theories the corresponding
Slavnov Taylor identities should be imposed as constraint  equations. The reason why
gauge invariance is broken  when the BPHZ method is applied to nonabelian gauge
theories lies in the subtraction process which is  based on expanding around an
external momentum and  thus modifying the structure of the corresponding
amplitude. In IR the amplitude is never modified. Moreover some modifications in the BPHZ framework  (Soft BPHZ Scheme) \cite{SOFT} must be introduced to handle infrared  divergencies because in the original formulation the  subtraction
is constructed at zero external momentum.

IR has been  applied  to various quantum field theoretical models including
dimension specific theories in which DR fails. For quantum
eletrodynamics, theories involving parity violating objects (Chern-Simons,
Chiral Schwinger Model), see \cite{PRD1}. For the study of  CPT
violation in an extended chiral version of quantum electrodynamics see
\cite{PRD2}. A comparison between IR, dimensional regularization, differential
renormalization and BPHZ  forest formula can be found in \cite{PRD3}. A constrained version of IR in which
certain arbitrary parameters may be fixed ab initio to render the theory gauge invariant throughout the calculation was built in \cite{ELOY} where we study the renormalization of QCD to one loop order. It was verified that the renormalization constants, defined minimally in terms of basic divergent integrals, satisfy the correct relations imposed by the Slavnov-Taylor identities (the finite part is gauge invariant because the amplitude is not modified in IR). In \cite{NLOOP} a model calculation using $\phi^3$ theory in $6$
dimensions illustrates how IR works when overlapping divergencies occur.  In \cite{JHEP}
it is  shown that IR  can be made manisfestly supersymmetric invariant.
This is illustrated  by renormalizing the massless Wess-Zumino model  and
calculating the beta function to three loop order. Phenomenological applications to a Gauged
Nambu-Jona Lasinio model can be found in \cite{CO} and to the linear sigma model in \cite{DIAS}.

In this contribution our goal is twofold: to motivate IR as the ideal arena
to parametrize  arbitrary quantities which are not fixed on renormalization
group grounds being particularly suitable for the description of anomalies and study two dimensional gravitational anomalies  for  Weyl massless fermions  immersed in a gravitational background using IR and carrying arbitrary quantities till the end of the calculation to be fixed on physical grounds allowing for a democratic discussion about the anomalous Ward identities once they proceed from the same amplitude. This approach goes back to John Bell's conception \cite{JACKBELL} that in the triangle anomaly, for instance, the amplitude containing a singular integral exhibits an arbitrariness in the calculation which should be fixed only when some external information is added. In \cite{JACKIW}, Jackiw shows that for superficially divergent amplitudes it can happen that  radiative corrections are finite and not determined by (the symmetry content of) the theory in which case, just as for infinite corrections, their values should be fixed by the experiment. Most regularization frameworks fail in this sense by assigning a definite value for such arbitrariness.

We conclude that: 1) the Einstein and Weyl Ward identities may be reduced to a set of algebraic equations for the arbitrary parameters; 2) whilst for the chiral Schwinger model and for the triangle anomaly different values of such parameters shift the anomaly between the axial and vector sectors, the axial part of the Einstein anomaly is always violated;  3) such breaking is  based on the interplay of arbitrary parameters in a way which clearly does not depend on the renormalization or subtraction procedure; 4) putting the Weyl and Einstein anomalies on equal footing we end up with two arbitrary regularization dependent parameters in accordance with Smailagic and Spalucci \cite{SMAILAGIC} who adopted the Fujikawa approach  and 5) keeping the renormalization scheme independence we may as well choose to preserve the Weyl Ward identities and violate the pure tensor part of the Einstein Ward identity, contrarily to what happens in DR which is known to preserve tranversality in the vector sector.

The paper is organized as follows: in section \ref{SQED} we review briefly the role of arbitrary parameters of IR in connection with gauge and momentum routing invariance by studying the vacuum polarization tensor of quantum electrodynamics. In section \ref{SCHI} we show IR at work by evenly displaying the chiral anomalies using as example the triangle anomaly and the (chiral) Schwinger model. Finally we apply such procedure in the case of two dimensional gravitational anomalies in section \ref{SGA}.

\section{Gauge invariance, momentum routing and arbitrary quantities}
\label{SQED}

In order to illustrate the modus operandi of IR as well as show the relation between arbitrary (regularization dependent) parameters, gauge invariance and momentum routing in a one loop Feynman diagram,
consider the vacuum polarization tensor of QED (see \cite{PRD2} for a detailed discussion).
To one loop order with $p$ being the external momentum it reads:
\be
\Pi_{\mu \nu} (p^2) = -e^2 \ikl tr \{ \gamma_\mu S(k+p) \gamma_\nu (k) \} \, ,
\label{vpt}
\ee
where $\int_k \equiv \int d^4k/(2 \pi)^4$ and $S$ is the free fermion propagator. After taking the Dirac trace we separate the ultraviolet divergent content of the amplitude as  loop integrals which depend only on the internal momenta through the following algebraic identity applied at the level of the integrand
\bq
&& \frac{1}{[(k+p)^2-m^2]} = \sum_{j=0}^{N} \frac{(-1)^j (p^2+ 2 p
\cdot k)^j}{(k^2-m^2)^{j+1}} \nonumber \\ &&+
\frac{(-1)^{N+1} (p^2 + 2 p\cdot k)^{N+1}}{(k^2 -m^2)^{N+1}
[(k+p)^2-m^2]} \, ,
\label{eqn:rr}
\eq
to obtain
\bq
&&\frac{\Pi_{\mu \nu }}{4} = 2 \ikl \frac{k_\mu k_\nu}{(k^2-m^2)^2}- g_{\mu \nu} \ikl \frac{k^2}{(k^2-m^2)^2} \nonumber \\ && + m^2 g_{\mu \nu} \ikl \frac{1}{(k^2-m^2)^2} - p^2
\ikl \frac{2 k_\mu k_\nu}{(k^2-m^2)^3} \nonumber \\ && + 8 p^\alpha p^\beta \ikl \frac{k_\mu k_\nu k_\alpha k_\beta}{(k^2-m^2)^4} - 2 p^\alpha p_\nu \ikl \frac{k_\alpha k_\mu}{(k^2-m^2)^3}\nonumber \\
&& - 2 p^\alpha p_\mu \ikl \frac{k_\alpha k_\nu}{(k^2-m^2)^3} - p^2 g_{\mu \nu} \ikl \frac{k^2}{(k^2-m^2)^3}- \nonumber \\ && 4 g_{\mu \nu} p_\alpha p_\beta \ikl \frac{k^2 k_\alpha k_\beta}{(k^2-m^2)^4} +
2 g_{\mu \nu} p_\alpha p_\beta \ikl \frac{k_\alpha k_\beta}{(k^2-m^2)^3} \nonumber \\ &&  -\frac{b}{3} \Big( p^2g_{\mu
\nu }-p_\mu p_\nu \Big)\Bigg( \frac{1}{3}+\frac{(2m^2+ p^2)}{p^2}
Z_0(p^2;m^2)\Bigg)
\label{qedinf}
\eq
where
\be b \equiv \frac{i}{(4 \pi)^2} \ee
and $Z_0 (p^2;m^2) = \int_0^1 dz \, \ln [(p^2 z(z-1)+m^2)/(m^2) ]$ is finite.
At one loop order the ultraviolet divergent behaviour can be written solely as
\bq
I_{log} (m^2) &=& \int \frac{d^4k}{(2 \pi)^4} \frac{1}{(k^2-m^2)^2} \,\, \mbox{and} \nonumber \\
I_{quad} (m^2) &=& \int \frac{d^4k}{(2 \pi)^4} \frac{1}{(k^2-m^2)} \, ,\mbox{etc.} ,
\eq
(similar basic divergent integrals appear at higher loop order \cite{JHEP}) taking into account that the following differences between divergent integrals of the same degree of divergence are finite and regularization dependent
\be
\Upsilon_{\mu \nu}^2 \equiv   g_{\mu \nu} I_{quad} (m^2) - 2 \Theta_{\mu
\nu}^{(2)}   = \alpha_1  g_{\mu \nu} \, ,
\label{CR4Q1}
\ee
\be
\Upsilon_{\mu \nu}^0 \equiv g_{\mu\nu} I_{log} (m^2) - 4 \Theta_{\mu \nu}^{(0)}
= \alpha_2 g_{\mu \nu}
 \label{CR4L1}
\ee
\be
\Upsilon_{\mu \nu \alpha \beta}^2 \equiv
g_{\{\mu\nu}g_{\alpha\beta \}} I_{quad} (m^2)  - 8 \Theta_{\mu \nu \alpha
\beta}^{(2)}   = \alpha_3 g_{\{\mu\nu}g_{\alpha\beta \}} \, ,
\label{CR4Q2}
\ee
\be
\Upsilon_{\mu \nu \alpha \beta}^0 \equiv
g_{\{\mu\nu}g_{\alpha\beta \}}  I_{log} (m^2)  - 24   \Theta_{\mu \nu \alpha
\beta}^{(0)}   =  \alpha_4 g_{\{\mu\nu}g_{\alpha\beta \}} \,
\label{CR4L2}
\ee
where
\begin{eqnarray}
\Theta_{\mu \nu}^{(0)}(m^2)&=&\int_k  \frac{k_\mu k_\nu}{(k^2-m^2)^3} \,\, ,
\nonumber \\ \Theta^{(2)}_{\mu \nu} (m^2)&=& \int_k \frac{k_\mu
k_\nu}{(k^2-m^2)^2} \,\, ,\nonumber \\ \Theta_{\mu \nu \alpha \beta}^{(0)} (m^2)
&=& \int_k \frac{k_\mu k_\nu k_\alpha k_\beta}{(k^2-m^2)^4}
\,\, , \nonumber \\ \Theta_{\mu \nu \alpha \beta}^{(2)} (m^2) &=&
\int_k \frac{k_\mu k_\nu k_\alpha k_\beta}{(k^2-m^2)^3}
\, ,
\end{eqnarray}
$g_{\{\mu\nu}g_{\alpha\beta \}}$ stands for
$g_{\mu\nu}g_{\alpha\beta}+g_{\mu\alpha}g_{\nu\beta}+g_{\mu\beta}g_{\nu\alpha}$,
and the   $\alpha_i$'s are arbitrary. Hence
(\ref{qedinf}) reduces to
\bq
\Pi_{\mu \nu } &=&  \widetilde{\Pi}_{\mu \nu}  + 4\Bigg(\Upsilon^2_{\mu\nu}-\frac{1}{2}p^2 \Upsilon^0_{\mu \nu}
+\frac{1}{3} p^{\alpha}p^{\beta}
\Upsilon^0_{\mu \nu \alpha
\beta} \nonumber \\ &-&
p^{\alpha}p_{\mu}\Upsilon^0_{\nu
\alpha}
-\frac{1}{2}p^{\alpha}p^{\beta}g_{\mu \nu}
\Upsilon^0_{\alpha \beta}  \Bigg)  \nonumber \\
&=& \widetilde{\Pi}_{\mu \nu} + (\alpha_1' m^2 g_{\mu \nu}+ \alpha_2'p^2 g_{\mu \nu} + \alpha_3'p_\mu p_\nu) \,\,\, {\mbox{with}} \nonumber \\
\widetilde{\Pi}_{\mu \nu} &=& \frac{4}{3} ( p^2 g_{\mu \nu }-p_\mu p_\nu )\Bigg( I_{log} (m^2)- b \Big(\frac{1}{3}+\frac{(2m^2+ p^2)}{p^2} \times \nonumber \\ &\times& Z_0(p^2;m^2)\Big) \Bigg)\, ,
\label{qedvpt2}
\eq
where we used (\ref{CR4Q1}) - (\ref{CR4L2}). In order to ensure transversality (gauge invariance) in (\ref{qedvpt2}) we must set $\alpha_i' = 0, i=1, 2, 3$. Some comments are in order. In \cite{PRD1}, \cite{PRD2} we showed that setting $\alpha_i's = 0 $ in  (\ref{CR4Q1})-(\ref{CR4L2}) implies that the one loop amplitude is momentum routing invariant and thus a shift in the  momentum integration variable is allowed. Indeed had we written (\ref{vpt}) as $-e^2 \int_k tr \{  \gamma_\mu S(k+k_1) \gamma_\nu S(k+k_2) \gamma_\nu\}$
with $k_1-k_2=p$ we would have obtained
\bq
&&\Pi_{\mu \nu } = \widetilde\Pi_{\mu \nu }(k_1-k_2) +
4\Bigg(\Upsilon^2_{\mu\nu}-\frac{1}{2}(k_1^2+k_2^2)\Upsilon^0_{\mu
\nu}
+ \nonumber \\ && + \frac{1}{3}(k_{1}^{\alpha}k_{1}^{\beta}+k_{2}^{\alpha}k_{2}^{\beta}
+k_{1}^{\alpha}k_{2}^{\beta})
\Upsilon^0_{\mu \nu \alpha
\beta} -
(k_1+k_2)^{\alpha}(k_1+k_2)_{\mu} \nonumber \\  && \Upsilon^0_{\nu
\alpha}  - \frac{1}{2}(k_1^{\alpha}k_1^{\beta}+k_2^{\alpha}k_2^{\beta})g_{\mu \nu}
\Upsilon^0_{\alpha \beta}  \Bigg)\,\, .
\label{qedvp}
\eq
showing clearly that the terms multiplying the $\Upsilon's$ are momentum routing dependent. DR, for instance, evaluates (\ref{CR4Q1})-(\ref{CR4L2}) to zero showing that it explicitly preserves gauge invariance. On the other hand perturbation theory may present some oddities such as preserving gauge invariance at the expense of adopting a special momentum routing \cite{CURRENT} e.g. in the AVV triangle anomaly. Amplitudes which manifest this feature usually have one axial vertex and should not be treated with naive dimensional regularization: whilst a shift in the integration variable is always possible in  DR, the algebraic properties of $\gamma_5$ matrix clash with analytical continuation on the space- time dimension. In fact a constrained version of IR in which all $\Upsilon's$ are set to vanish has been shown to yield gauge invariant amplitudes from the start. Such abbreviated version of IR fixes arbitrary local terms {\it {a priori}}
in such a way that the Ward and Slavnov-Taylor identities are directly fulfilled in a similar fashion as constrained DfR
\cite{DF}. In \cite{ELOY} we show that constrained IR is consistent with non-abelian gauge invariance by studying the Slavnov-Taylor identities to one loop order in quantum chromodynamics. To renormalize the field through the renormalization constant $Z_3$, $A^\mu = Z_3^{1/2} A^\mu_R $ we define $\Pi_{\mu \nu}^R = \Pi_{\mu \nu} + (p_\mu p_\nu -p^2 g_{\mu \nu}) (Z_3 -1)$. A minimal, mass independent within IR in comparison with DR and DfR \cite{PRD3} is defined through the identity in the four dimensional space-time
\be
I_{log} (m^2) = I_{log} (\lambda^2) + \frac{i}{(4 \pi)^2} \ln \Big( \frac{\lambda^2}{ m^2} \Big) \, ,
\label{re}
\ee
where $\lambda^2$ is an arbitrary constant which parametrizes the freedom in separating the divergent from the finite part and plays the role of renormalization group scale in IR. For a massless theory we may always introduce an infrared cutoff $\mu$ at the level of the propagators to define $I_{log} (\mu^2)$. As $\mu \rightarrow 0$ a genuine counterterm can be defined through (\ref{re}) for $\lambda^2 \ne 0$. For infrared safe models the $\ln \mu^2$ in (\ref{re}) will always cancel out with a similar term coming from the ultraviolet finite part of the amplitude. For instance taking
$m^2 = \mu^2 \rightarrow 0$ into (\ref{qedvpt2}) and using (\ref{re})yields
\be
\widetilde{\Pi}_{\mu \nu} = \frac{4}{3} ( p^2 g_{\mu \nu }-p_\mu p_\nu ) \Bigg( I_{log} (\lambda^2) + \nonumber \\
b \ln \Big( \frac{\lambda^2 \mbox{e}^2}{-p^2} \Big) - \frac{b}{3} \Bigg)\, ,
\ee
where e is the Euler number and $\lambda^2 \ne 0$, which in our minimal subtraction scheme gives $Z_3 = 1 + 4/3 \, i  I_{log} (\lambda^2)$. Finally using that $\partial I_{log} (\lambda^2)/\partial \lambda^2 = -b/\lambda^2$ allows us to obtain the $\beta$-function to this order namely $\beta = e^3 /(12 \pi^2)$. This procedure generalizes straightforwardly to higher loop orders. In \cite{JHEP} we calculate the three loop $\beta$ function for the (massless, supersymmetric) Wess-Zumino model. Divergences other than logarithmic may play an important role as well. In \cite{ELOY} we show that the quadratic divergences originated from the tadpoles are important to cancel out other quadratic divergences in QCD at one loop order in order to maintain gauge invariance.

\section{The Adler-Bardeen-Bell-Jackiw triangle and the chiral Schwinger model anomalies}
\label{SCHI}

In order to gain some insight into the study of two dimensional gravitational anomalies we briefly  restate the discussion on the AVV and chiral Schwinger model anomalies within IR \cite{PRD1},\cite{PRD2}. We emphasize on the origin of ambiguities and free parameters as well as on the importance of displaying the anomaly evenly amid the Ward identities which stem from a  Feynman graph unless Physics says otherwise. Such feature provides a sort of ``acid test" for regularizations and we conclude that IR is a good arena to study anomalies from the Feynman diagram viewpoint. The AVV triangle with arbitrary momentum routing reads
\bq
&& T^{AVV}_{\mu \nu \alpha} = -\int_k
\mbox{tr} \Big\{ \gamma_{\mu}
(k\hspace{-2.2mm}/
+ k_1\hspace{-2.2mm}/
 -m )^{-1} \gamma_{\nu}
(k\hspace{-2.2mm}/
+ k_2\hspace{-2.2mm}/
 -m )^{-1} \times \nonumber \\ && \gamma_{\alpha}
\times \gamma_{5}
(k\hspace{-2.2mm}/
+ k_3\hspace{-2.2mm}/
 -m )^{-1} \Big\}
+ {\stackrel{crossed}{term}}.
\eq
where the $k_i$'s are related to the external momenta $p$, $q$ and $p+q$ such that $k_2 - k_3 = p+q$
$k_1 - k_3 = p$ and $k_2 - k_1 =1 q$. Hence we  may parametrize the $k_i$'s as
\bq
k_1&=&\alpha p+ (\beta -1) q ,\nonumber \\
k_2&=&\alpha p+\beta q, \nonumber \\
k_3&=&(\alpha -1)p + (\beta -1) q,
\label{park}
\eq
for general $\alpha$ and $\beta$. Using the IR framework allows us to write \cite{PRD2}
\bq
T_{\mu \nu \alpha}^{AVV} &=& \tilde {T}_{\mu \nu \alpha}^{AVV}(p,q) \nonumber \\ &+& 4i \alpha_1
(\alpha-\beta +1)
 \epsilon_{\mu \nu \alpha \beta}(p-q)^{\beta}.
\eq
where we have set $\Upsilon_{\mu \nu}^0 \equiv \alpha_1 g_{\mu \nu}$ as in eqn. (\ref{CR4L1}) and $\tilde {T}_{\mu \nu \alpha}^{AVV}(p,q)$ is a function of the external momenta free of arbitrary parameters which satisfies in the zero mass limit \cite{PRD2}
\bq
p^\mu\tilde {T}_{\mu \nu \alpha}^{AVV}&=&-\frac
1{4\pi^2}\epsilon_{\mu \nu \alpha \beta}
p^\mu q^\beta \\
q^\nu\tilde{T}_{\mu \nu \alpha}^{AVV}&=&\frac
1{4\pi^2}\epsilon_{\mu \nu \alpha \beta}
p^\beta q^\nu \\
(p+q)^{\alpha}\tilde{T}_{\mu \nu \alpha}^{AVV}&=& 0 .
\eq
Thus the Ward identities read
\bq
p^{\mu}T_{\mu \nu \alpha}^{AVV}&=& \left\{-\frac 1{4\pi^2} -
4i\alpha_1
(\alpha -\beta +1)\right\} \epsilon_{\mu \nu \alpha \beta}p^{\mu}q^{\beta} \nonumber \\
q^{\nu}T_{\mu \nu \alpha}^{AVV}&=&\left\{\frac 1{4\pi^2}+
4i\alpha_1
(\alpha -\beta +1 )\right\}
\epsilon_{\mu \nu \alpha \beta}q^{\nu}p^{\beta} \nonumber \\
(p+q)^{\alpha}T_{\mu \nu \alpha}^{AVV}&=& -8i\alpha_1 (\alpha -\beta +1)
\epsilon_{\mu \nu \alpha \beta}p^{\alpha}q^{\beta}.
\label{wi}
\eq
which again illustrates the connection between the $\Upsilon$'s and arbitrary momentum routing. A redefinition of variables $\alpha_1 (\alpha -\beta +1) = i(1-a)/(32 \pi^2)$, for arbitrary $a$ yields
\bq
p^{\mu}T_{\mu \nu \alpha}^{AVV}&=&-\frac 1{8\pi^2}(1 + a)
\epsilon_{\mu \nu \alpha
\rho}p^{\mu}q^{\rho} \\
(p+q)^{\alpha}T_{\mu \nu \alpha}&=& \frac 1{4\pi^2}(1 - a)
\epsilon_{\mu \nu \alpha
\rho}p^{\alpha}q^{\rho}.
\eq
which evidently show that the anomaly  floats between the axial and vector channels and therefore the correct answer
for the triangle graph is not intrinsic to it. Finally let us turn our attention  to the Schwinger model and its chiral version to explore the analogy with fermions in a two dimensional curved background. Hereafter we work in the $1+1$ dimensional space-time and
$\int_k$ will always stand for $\int \, d^2 k/(2 \pi)^2$. In $1+1$ dimensions the following differences between superficially logarithmically divergent integrals are finite and regularization dependent (analogous to relations (\ref{CR4Q1})-(\ref{CR4L2})):
\bq
\Xi^{0}_{\mu \nu} = \int_k \frac{g_{\mu \nu}}{\dena} -
2 \int_k \frac{k_\mu k_\nu}{\dena^2} \, ,\nonumber \\
\Xi^{0}_{\mu \nu \sigma \rho} = \int_k
\frac{g_{\{ \mu\nu} g_{\sigma \rho \}}}{\dena} - 8 \int_k \frac{k_\mu k_\nu k_\sigma
k_\rho}{\dena^3}  \, , \nonumber \\
\Xi^{0}_{\alpha \beta \mu \nu \sigma \rho} = \int_k
\frac{g_{\{ \alpha \beta}g_{\mu \nu}g_{\rho \sigma \}}}{\dena} - 48 \int_k \frac{k_\mu k_\nu k_\alpha k_\beta k_\sigma k_\rho}{\dena^4}  \,
\label{CR2D}
\eq
which we write as $\Xi^{0}_{\mu \nu}= \alpha_1 g_{\mu \nu}$,   $\Xi^{0}_{\mu \nu \sigma \rho}=\alpha_2 g_{\{ \mu\nu} g_{\sigma \rho \}}$ and $\Xi^{0}_{\alpha \beta \mu \nu \sigma \rho}= \alpha_3 g_{\{ \alpha \beta}g_{\mu \nu}g_{\rho \sigma \}}$ with $\alpha_i$ arbitrary and the curly brackets standing for a symmetrization on the Lorentz indices.

In the Schwinger model, the massless photon of the tree
approximation acquires the mass $e^2/\pi$, $e$ is the coupling
constant, at the one loop level (which is exact in this case). The
mass generation is seen at order $A^2$ of the gauge potential so we
need to compute the vacuum polarization tensor \be \Pi_{S}^{\mu \nu}
(p) = i \mbox{tr} \int \dd \, \gamma^\mu
\frac{i}{k\hspace{-2mm}/}\gamma^\nu
\frac{i}{k\hspace{-2mm}/+p\hspace{-2mm}/} \, . \label{pmnir} \ee
Adopting the IR framework we obtain \be \Pi_{S}^{\mu \nu} (p) =
\frac{1}{\pi} \Big(\frac{\alpha_1 + 2}{2}g^{\mu \nu} - \frac{p^\mu
p^\nu}{p^2}\Big) \, , \label{pmnsm} \ee where we have used
(\ref{CR2D}). The  choice $\alpha_1 = 0$ can be used in
(\ref{pmnsm}) to ensure gauge invariance. It plays the role of an
undetermined local part in the quadratic term of the effective
action. In other words we could say that we had to choose $\alpha_1$
to vanish in order to explain the photon mass $m^2 = e^2/\pi$ should
this model be real. In the chiral Schwinger model we substitute the
vector interaction with a chiral interaction namely
$A\hspace{-2mm}/ \rightarrow  (1+\gamma_5)A\hspace{-2mm}/$, where
$\gamma_5 = \gamma_0 \gamma_1$ satisfies $\gamma_5 \gamma_\mu =
\epsilon^{\mu \nu} \gamma_\nu $ which allows us to write \bq
\Pi_{\chi }^{\mu \nu} (p) &=& \Pi_{S}^{\mu \nu} (p) + g_{\alpha
\beta} \Big( \epsilon^{\nu \alpha}\Pi_{S}^{\mu \beta} (p) +
\epsilon^{\mu
\alpha}\Pi_{S}^{\beta \nu} (p) \Big) +\nonumber \\
&+&
\epsilon^{\mu
\alpha}\epsilon^{\nu \beta}\Pi_{S \,\, \alpha \beta}
(p) \, .
\label{pmn-cscm}
\eq
 An analogous calculation within IR  leads us to the result
\be \Pi_{\chi }^{\mu \nu} (p) = \frac{1}{\pi} \Bigg( (\alpha_1 +
2)g^{\mu \nu} -  (g^{\mu \alpha}+\epsilon^{\mu \alpha})
\frac{p_\alpha p_\beta}{p^2} (g^{\beta \nu}- \epsilon^{\beta \nu})
\Bigg) \, . \ee Unlike the Schwinger model, imposing gauge
invariance does not fix the value of $\alpha_1$ since \be p_\mu
\Pi_{\chi }^{\mu \nu} (p) = \frac{1}{\pi}\Big( (\alpha_1 + 1) p^\nu
- \tilde{p}^\nu \Big) \, , \ee $ \tilde{p}^\nu=\epsilon^{\nu
\alpha}p_\alpha$ which shows that the longitudinal part does not
vanish for any value of $\alpha_1$. This is a manifestation of the
anomalous non-simultaneous conservation of the chiral and vector
current since \bq p_\nu {\Pi}^{\mu \nu}_{5} &=& - \frac{\alpha_1 +
2}{2
\pi} \,\, \tilde{p}^\mu \,\,\, ,\mbox{whereas} \nonumber \\
p_\nu {\Pi}^{\mu \nu}_{S} &=& \frac{p^\mu}{2 \pi} \alpha_1
\label{avwi} \eq where $ {\Pi}^{\mu \nu}_{5} = \epsilon^{\nu
\kappa}( \Pi^\mu_\kappa)_S$. The anomaly of magnitude $-1/\pi$
floats between the axial and vector Ward identities through one
parameter, $\alpha_1$.

The chiral Schwinger model can be exactly solved to find that for
$\alpha_1 > - 1$ it is a unitary and positive definite model, in
which the photon acquires a mass \be m^2 = \frac{e^2}{\pi}
\frac{(\alpha_1 +2)^2}{\alpha_1 + 1} \ee An equivalent formulation
in a bosonized version of the model places $\alpha_1$ as arising
from ambiguities in the bosonization procedure \cite{PRD1}. This
time theory's unitarity and positivity constraints only establish a
range of values for the arbitrary parameter, namely $\alpha_1 > -1$
.

\section{Two dimensional gravitational anomalies}
\label{SGA}

Consider chiral (Weyl) fermions coupled to gravitation as an external nonquantized field in the $1+1$ dimensional space time. Gravitation considered as a gauge theory may exhibit anomalies expressed by the breakdown of general coordinate transformation invariance (Einstein anomaly), rotation invariance in the tangent frame (Lorentz anomaly) and conformal symmetry (Weyl anomaly). Such breakings manifest themselves in the energy momentum tensor as a violation in its classical conservation law, the existence of  an antisymmetric part and a nonvanishing trace, respectively. The seminal work on gravitational anomalies was written by Alvarez-Gaum\'e and Witten \cite{GAUME} followed by Langouche \cite{LANGOUCHE} and Tomiya \cite{TOMIYA} within perturbation theory. Also other approaches such as  heat kernel method \cite{LEUTWYLER}, Fujikawa's method \cite{FUJIKAWA}, \cite{SMAILAGIC} and differential geometry and topology formulation  (see \cite{BERTLMANN-B} for a review) can be used to display such anomalies.

Here we will be studying the interplay between  Einstein and Weyl anomalies since Lorentz and Einstein anomalies are not independent  as stated by the Bardeen-Zumino theorem \cite{ZUMINO}. Thus our quantized energy momentum tensor is symmetric from the start. The Lagrangian reads
\be
{\cal{L}} = i e E^{a \mu} \overline{\psi}
\gamma_a \frac{1}{2} \overleftrightarrow{\partial_\mu} \frac{1 \pm
\gamma_5}{2} \psi,
\ee
where $e_{~ \mu}^{a}$ is the zweibein,
$E_{a}^{~ \mu}$ its inverse and $e=det ~ (e^{a}_{~ \mu})$. There is no need to define a spin connection through a covariant derivative in $1+1$ dimensions what greatly simplifies the calculation \cite{BERTLMANN}. Our conventions are $\eta_{\mu \nu} = {\mbox{diagonal}} (1,-1)$, $\varepsilon^{01}=1=-\varepsilon^{10}$, $\gamma^0 = \sigma^2$, $\gamma^1 = i \sigma^1$, $\gamma_5=\gamma^0 \gamma^1=\sigma^3$. It is sufficient for our purposes to linearize the gravitational field $g_{\mu \nu}=\eta_{\mu \nu} + \kappa h_{\mu \nu} +
\mathcal{O}(\kappa^2)$ and  $e^{a}_{~ \mu}=\eta^{a}_{~ \mu}+\frac{1}{2} \kappa h^{a}_{~
\mu}+\mathcal{O}(\kappa^2)$ to obtain the interaction Lagrangian
\be
{\cal{L}}^{lin}_{I} = - \frac{i}{4} \Bigg{(} h^{a
\mu}\overline{\psi}\gamma_a\frac{1 \pm
\gamma_5}{2}\overleftrightarrow{\partial^{\psi}_{\mu}}\psi +
h^{\mu}_{~ \mu} \overline{\psi}\gamma^a\frac{1 \pm
\gamma_5}{2}\overleftrightarrow{\partial^{\psi}_{a}}\psi \Bigg{)}.
\ee
in which $\partial^{\psi}_{a}$ acts only on the spinor. The energy momentum tensor is defined as ${\cal{L}}^{lin}_{I} = - 1/2 \,\, h_{\mu \nu} T^{\mu \nu}$. The quantum correction comes from the two point function
\be
\label{2pontos} T_{\mu \nu \sigma \rho}(p) = i \int d^2 x \, e^{i p
x} \langle 0 | T [T_{\mu \nu}(x)T_{\rho \sigma}(0)] | 0\rangle \, ,
\ee
which in momentum space corresponds to a fermion loop contribution to the graviton propagator (external current). It reads
\bq
&&
T_{\mu \nu
\rho \sigma}(p) = - \frac{i}{16} \mbox{tr} \int_k \Bigg{(}
\Big[ \gamma_\mu (p+2 k)_\nu + \gamma_\nu (p+2 k)_\mu \Big]  \nonumber \\  &&
{\cal{P}}_{\pm} \frac{p\hspace{-2.2mm}/+ k\hspace{-2.2mm}/+m}{(p+k)^2-\mu^2} \Big[\gamma_\rho (p+2
k)_\sigma + \gamma_\sigma (p+2 k)_\rho \Big] \nonumber \\&&
{\cal{P}}_{\pm} \frac{k\hspace{-2.2mm}/ + m}{k^2-\mu^2} \Bigg{)} \, ,
\label{ampl}
\eq
where ${\cal{P}}_{\pm}= (1 \pm \gamma_5)/2$ and $\mu$ is a fictitious mass which we set to zero in the end of the calculation. We follow the notation and conventions established by Bertlmann in \cite{BERTLMANN} where is shown an equivalence between the dispersive approach and dimensional regularization. Lorentz covariance enable us to separate $T_{\mu \nu \rho \sigma}$ into a pure tensor and a pseudo-tensor part $T_{\mu \nu \sigma \rho} = T_{\mu \nu \sigma \rho}^{V}+T_{\mu \nu \sigma \rho}^{A}$ which may be written in terms of form factors $T_i(p^2)$,

\bq && T_{\mu \nu \sigma \rho}^{V} = p_\mu p_\nu p_\sigma p_\rho
T_1(p^2) + (p_\mu p_\nu g_{\rho \sigma}+p_\rho p_\sigma g_{\mu
\nu})T_2(p^2) \nonumber \\ && +(p_\mu p_\rho g_{\sigma
\nu}+p_\mu p_\sigma g_{\rho \nu}+p_\rho p_\nu g_{\mu \sigma}+p_\nu
p_\sigma g_{\mu \rho})T_3(p^2)+ \nonumber \\ && + g_{\mu
\nu}g_{\rho \sigma}T_4(p^2)+(g_{\mu \rho}g_{\nu \sigma}+g_{\mu
\sigma}g_{\rho \nu})T_5(p^2) \eq

\bq && T_{\mu \nu \sigma \rho}^{A}(p) = (\varepsilon_{\mu \tau}
p^\tau p_\nu p_\rho p_\sigma+\varepsilon_{\nu \tau} p^\tau p_\mu
p_\rho p_\sigma+\varepsilon_{\rho \tau} p^\tau p_\nu p_\nu
p_\sigma \nonumber \\ && + \varepsilon_{\sigma \tau} p^\tau p_\mu p_\nu
p_\rho)T_6(p^2) + (\varepsilon_{\mu \tau} p^\tau
p_\nu g_{\rho \sigma}+\varepsilon_{\nu \tau} p^\tau p_\mu g_{\rho
\sigma}+ \nonumber \\ && \varepsilon_{\rho \tau} p^\tau p_\sigma g_{\mu
\nu}+\varepsilon_{\sigma \tau} p^\tau p_\rho g_{\mu \nu})T_7(p^2)
 + \Big{[}\varepsilon_{\mu \tau} p^\tau (p_\rho
g_{\nu \sigma}+ \nonumber \\ && p_\sigma g_{\nu \rho})+ \varepsilon_{\nu \tau}
p^\tau (p_\rho g_{\mu \sigma}+p_\sigma g_{\mu \rho})+\varepsilon_{\rho \tau} p^\tau (p_\mu
g_{\nu \sigma}+p_\nu g_{\mu \sigma}) \nonumber \\ && + \varepsilon_{\sigma \tau}
p^\tau (p_\mu g_{\nu \rho}+p_\nu g_{\mu \rho}) \Big{]} T_8(p^2).
\eq
We assume the quantized energy momentum tensor to be symmetric, $T_{\mu \nu \sigma \rho}=T_{\nu \mu \sigma \rho}$ (no violation of Lorentz symmetry) \cite{ZUMINO}.  Thus the Einstein and Weyl Ward identities (EWI and WWI), $p^\mu T_{\mu \nu \rho \sigma}(p) = 0$ and $g^{\mu \nu}T_{\mu \nu \rho \sigma}(p) = 0$, may be cast as follows
\be
\mbox{Pure Tensor Part EWI}  \left\{ \begin{array}{ll}
         p^2 T_1 + T_2 + 2 T_3 & = 0\\
         p^2 T_2 + T_4 & = 0 \\
     p^2 T_3 + T_5 & = 0
     \end{array} \right.
     \label{wi1}
\ee
\be
\mbox{Axial Part EWI}  \left\{ \begin{array}{ll}
        3 p^2 T_6 + 2 T_7 + 4 T_8 & = 0\\
         p^2 T_7 + 2 p^2 T_8 & = 0
     \end{array} \right.
     \label{wi2}
\ee
\be
\mbox{Pure Tensor Part WWI}  \left\{ \begin{array}{ll}
         p^2 T_1 + 2 T_2 + 4 T_3 & = 0\\
         p^2 T_2 + 2 T_4 + 2 T_5 & = 0
     \end{array} \right.
     \label{wi3}
\ee
\be
\mbox{Axial Part WWI}  \left\{ \begin{array}{ll}
        p^2 T_6 + 2 T_7 + 4 T_8 & = 0\\
        \end{array} \right.
    \label{wi4}
\ee
It is straightforward to see that after taking the Dirac trace in (\ref{ampl}) and using that in $1+1$ dimensions
$\gamma_\mu \gamma_5 = - \varepsilon_{\mu \nu} \gamma^\nu$, the amplitude's pure tensor and axial parts may be written in terms of
\be
\label{tni}
T_{\mu
\nu \rho \sigma}^{n i} = \int_k \frac{q_\nu q_\sigma r_\mu
k_\rho+q_\nu q_\sigma r_\rho k_\mu - (r \cdot k) q_\nu q_\sigma
g_{\rho \mu}}{[(k+p)^2-m^2][k^2-m^2]}
\ee
where $q=p+2 k$ and  $r=p+k$ as
\be
\label{vet} T_{\mu \nu \rho \sigma}^{V} = T_{\mu \nu \rho
\sigma}^{n i}+T_{\nu \mu \rho \sigma}^{n i}+T_{\mu \nu \sigma
\rho}^{n i}+T_{\nu \mu \sigma \rho}^{n i},
\ee
\bq T_{\mu \nu \rho \sigma}^{A} & = &\mp \frac{1}{2}\Big{\{}
 (\varepsilon_{\mu}^{\tau} T_{\tau \nu \rho \sigma}^{n
i}+\varepsilon_{\rho}^{\tau} T_{\mu \nu \tau \sigma}^{n
i})+ (\mu \leftrightarrow \nu)  \nonumber\\
& + &  (\sigma \leftrightarrow \rho) + (\mu \leftrightarrow \nu, \sigma \leftrightarrow \rho)
\Big{\}}.
\eq
Defining $I_\mu = \int_k k_\mu/{\cal{D}}$, $I_{\mu \nu}^{k^2}= \int_k (k^2 k_\mu k_\nu)/{\cal{D}}$, etc. with
${\cal{D}}=[(k+p)^2-m^2][k^2-m^2]$ yields
\begin{eqnarray}
&& T_{\mu \nu \sigma \rho}^{ni}=p_\nu p_\sigma p_\mu I_\rho+p_\nu
p_\sigma p_\rho I_\mu + 2 p_\nu p_\sigma I_{\rho \mu}+ 2 p_\nu p_\mu
I_{\rho \sigma}
\nonumber\\
&& + 4 p_\nu I_{\rho \sigma \mu}+ 2 p_\mu p_\sigma
I_{\rho \nu} + 4 p_\sigma I_{\rho \mu \nu}+ 4 p_\mu
I_{\rho \nu \sigma}+ \nonumber \\ && 8 I_{\rho \mu \nu \sigma}+2 p_\nu p_\rho
I_{\sigma \mu}+ 2 p_\rho p_\sigma I_{\mu \nu}+ 4 p_\rho I_{\nu \mu
\sigma} \nonumber \\ &&  - p^\alpha p_\nu p_\sigma g_{\rho \mu}
I_\alpha - I^{k^2} p_\nu p_\sigma g_{\rho \mu} - 2 p^\alpha p_\nu
g_{\rho \mu} I_{\alpha \sigma} - \nonumber \\ && 2 p_\nu g_{\rho \mu} I_\sigma^{k^2}
- 2 p^\alpha p_\sigma g_{\rho \mu} I_{\alpha \nu} -
2 p_\sigma g_{\rho \mu} I_\nu^{k^2} - \nonumber \\ && 4 p^\alpha g_{\rho \mu}
I_{\alpha \nu \sigma} - 4  g_{\rho \mu} I_{\nu
\sigma}^{k^2}\,.
\label{tni}
\end{eqnarray}
Now we evaluate $I_{\mu, \mu \nu, ...}$  within IR in order to isolate the divergencies as basic divergent integrals. For instance, using (\ref{CR2D}) we write
\bq
I_{\mu \nu} &=& \frac{g_{\mu \nu}}{2} I_{log}(\mu^2) + \alpha_1 g_{\mu \nu} \nonumber \\ &+& \tilde{b} ~ \Bigg{\{} \Bigg( g_{\mu \nu} - \frac{p_\mu p_\nu}{p^2} \Bigg)
\Bigg[ \ln\Big(\frac{\mu^2}{-p^2}\Big)+1\Bigg] \Bigg{\}}
\eq
where $\tilde{b}=i/(4 \pi)$ and $I_{log}(\mu^2) = \int_k 1/(k^2-\mu^2)$ can be shown to satisfy
\be
I_{log}(\mu^2) = I_{log}(\lambda^2) + \tilde{b} \ln \Big( \frac{\mu^2}{\lambda^2} \Big) \, .
\label{re2}
\ee
Notice the appearance of the arbitrary regularization dependent parameter $\alpha_1$. The evaluation of $I's$ in (\ref{tni}) in IR is straightforward. It  will give rise to the other $\alpha_i$'s in (\ref{CR2D}) which will enter in the expression for the form factors $T_i (p^2)$ each of which receives contributions from differents terms in (\ref{tni}).
Here we quote the results.
\bq
T_1 &=&\frac{1}{24 \pi p^2} \, , \\
\nonumber \\ \label{t2} T_2 &=&- \frac{1}{18 \pi} - \frac{1}{48 \pi}
\ln \Big( \frac{\mu^2}{-p^2}\Big) - \nonumber \\ &-& \frac{i}{16} \Big{[} \frac{4}{3} I_{log}(\mu^2) - 4
\alpha_1 + 8 \alpha_2 - \frac{16}{3} \alpha_3 \Big{]}.
\eq
Some comments are in order before we proceed. In (\ref{t2}) the fictitious mass $\mu$ vanishes. That makes
$I_{log}(\mu^2)$ itself an hybrid object since it is both ultraviolet and infrared divergent. In order to correctly display the ultraviolet behaviour of the theory we use (\ref{re2}) for $\lambda^2 \ne 0$. Then the infrared piece
coming from the ultraviolet finite part in $T_2$ will cancel out. We end up with
\bq
T_2 &=&- \frac{1}{18 \pi} - \frac{1}{48 \pi}
\ln \Big( \frac{\lambda^2}{-p^2}\Big) - \nonumber \\ &-& \frac{i}{16} \Big{[} \frac{4}{3} I_{log}(\lambda^2) - 4
\alpha_1 + 8 \alpha_2 - \frac{16}{3} \alpha_3 \Big{]}.
\eq
The arbitrary constant $\lambda^2$  parametrizes a renormalization scheme and plays the role of renormalization group scale in a renormalizable model. We have shown in \cite{PRD3}, \cite{JHEP} that the (minimal) subtraction of $I_{log}(\lambda^2)$ leaves an ultraviolet finite part as a function of $\lambda$ which satisfies a Callan-Symanzik renormalization group equation. Coincidently such finite part is identical to one obtained within differential renormalization up to an immaterial rescaling of $\lambda$ \cite{PRD3}. One can find similar relations to (\ref{re2})
valid at higher loop order including the case of overlapping divergencies \cite{JHEP}. The other form factors follow the same reasoning. They read:
\bq
T_3 &=& \frac{1}{144 \pi} + \frac{1}{96 \pi} \ln \Big( \frac{\lambda^2}{-p^2}\Big) + \nonumber \\
&-& \frac{i}{16} \Big{[} - \frac{2}{3} I_{log}(\lambda^2) - 5
\alpha_1 + 10 \alpha_2 - \frac{16}{3} \alpha_3 \Big{]}\, ,  \\
T_4 &=&\frac{p^2}{18 \pi} + \frac{p^2}{48 \pi} \ln \Big( \frac{\lambda^2}{-p^2}\Big) \nonumber \\
&-& i \frac{p^2}{16} \Big{[} - \frac{4}{3} I_{log}(\lambda^2) + 4 \alpha_2
- \frac{8}{3} \alpha_3 \Big{]} \, ,   \\
T_5 &=&- \frac{p^2}{144 \pi} - \frac{p^2}{96 \pi} \ln \Big( \frac{\lambda^2}{-p^2}\Big) \nonumber \\
&-& i \frac{p^2}{16}
\Big{[} \frac{2}{3} I_{log}(\lambda^2) - 4 \alpha_1 + 6 \alpha_2 -
\frac{8}{3} \alpha_3 \Big{]} \, , \\  T_6 &=& \pm
\Bigg{(} - \frac{1}{96 \pi p^2}  \Bigg{)} \, ,\\
T_7 &=& \pm \Bigg{(} \frac{1}{72 \pi} + \frac{1}{192 \pi} \ln \Big( \frac{\lambda^2}{-p^2}\Big) \nonumber \\ &+&
\frac{i}{64} \Big{[} \frac{4}{3} I_{log}(\lambda^2) - 4 \alpha_1 + 8
\alpha_2 - \frac{16}{3} \alpha_3 \Big{]} \Bigg{)} \, ,\\
T_8 &=& \pm \Bigg{(} -\frac{1}{576 \pi} - \frac{1}{384
\pi} \ln \Big( \frac{\lambda^2}{-p^2}\Big) \nonumber \\ &+& \frac{i}{64} \Big{[} - \frac{2}{3} I_{log}(\lambda^2) -
5 \alpha_1 + 10 \alpha_2 - \frac{16}{3} \alpha_3 \Big{]} \Bigg{)} .
\eq
Note that $T_1 = \mp 4 T_6$, $T_2 = \mp 4 T_7 = - (1/p^2) T_4 $ and $T_3 = \mp 4 T_8 = -(1/p^2) T_5$.

Finally we are ready to write the Ward identities (\ref{wi1})-(\ref{wi4}) using the above displayed form factors. A simple substitution yields
\begin{itemize}
\item Einstein Ward Identity:
\be
\mbox{Pure Tensor Part}  \left\{ \begin{array}{ll}
         -14 \alpha_1 + 28 \alpha_2 - 16 \alpha_3 & = 0\\
         - 4 \alpha_1 + 12 \alpha_2 - 8 \alpha_3 & = 0 \\
     -9 \alpha_1 + 16 \alpha_2 - 8 \alpha_3  & = 0
     \end{array} \right.
     \label{ewi1}
\ee
\be
\mbox{Axial Part}  \left\{ \begin{array}{ll}
        -28 \alpha_1 + 56 \alpha_2 - 32 \alpha_3  & = \frac{2}{3 \pi}\\
         -14 \alpha_1 + 28 \alpha_2 - 16 \alpha_3  & = - \frac{2}3 \pi{}
     \end{array} \right.
     \label{ewi2}
\ee
\item Weyl Ward Identity:
\be
\mbox{Pure Tensor Part}  \left\{ \begin{array}{ll}
         -28 \alpha_1 + 56 \alpha_2 - 32 \alpha_3  & = -\frac{2}{3 \pi}\\
         -12 \alpha_1 + 28 \alpha_2 - 16 \alpha_3  & = \frac{2}{3 \pi}
     \end{array} \right.
     \label{ewi3}
\ee
\be
\mbox{Axial Part}  \left\{ \begin{array}{ll}
        -28 \alpha_1 + 56 \alpha_2 - 32 \alpha_3  & = - \frac{2}{3 \pi}\\
        \end{array} \right.
    \label{ewi4}
\ee
\end{itemize}

It is important to observe that in calculating the Ward identities (\ref{ewi1})-(\ref{ewi4}), which may lead to anomalies as we shall discuss,
the terms  proportional to $\ln ( \frac{\lambda^2}{-p^2})$  present in the form factors as well as the ultraviolet divergencies  $I_{log}(\lambda^2)$ cancel out. Recall that as a particular value of $\lambda$ defines a renormalization scheme, our treatment in IR clearly does not depend on the subtraction point as far as the Ward identities as concerned.
This is obviously desired otherwise the study of gravitational anomalies from the effective Lagrangian  we used would be meaningless. On the other hand within DR the independence of the renormalization scheme comes at the expense of preserving the pure tensor part of the Einstein Ward identity \cite{BERTLMANN},\cite{TOMIYA}. As we shall see this is a built in feature of DR because it is taylored to preserve pure vector gauge symmetry, just as in the case of the chiral anomaly discussed in the previous section. In other words, since DR evaluates the arbitrary regularization dependent parameters $\alpha_i$ to zero, as one can easily show, the pure tensor Einstein Ward identities  (\ref{ewi1}) are automatically satisfied.

We follow Bell and Jackiw's \cite{JACKIW} ideas in the sense of maximally preserving the democracy of the anomaly among the Ward identities since they stem from the same amplitude. This   investigation requires the solution of the algebraic equations for the $\alpha_i$'s. Since (\ref{ewi4}) is contained in (\ref{ewi3}) the fulfillment of the pure tensor part of the WWI implies that its axial part is satisfied as well. That is to say, the Weyl Ward identities can be satisfied
for arbitrary $\alpha_3$, $\alpha_1 = 1/(2 \pi)$ and $\alpha_2=(5 + 12 \pi \alpha_3)/(21 \pi)$. As for the Einstein Ward identities (\ref{ewi1})-(\ref{ewi2}) while its pure tensor part  admits only the trivial solution $\alpha_1 = \alpha_2 = \alpha_3 = 0$ the, the axial part can never be satisfied which implies that there is an intrinsic anomaly in this sector, and therefore it is a genuine quantum effect. This conclusion is in agreement  with the dispersion relation treatment
\cite{BERTLMANN}. However following our treatment we may as well choose to preserve the Weyl Ward identities and violate the pure tensor part of the Einstein identity what has no room within DR. Finally, by considering the Einstein and Ward identities on equal footing, we are left with two arbitrary, regularization dependent parameters, in consonance with the treatment using Fujikawa method by Smailagic and Spalluci in \cite{SMAILAGIC}.


\end{document}